\documentclass[12pt]{article}
\usepackage{amssymb}

\makeatletter
\def\numberbysection{\@addtoreset{equation}{section}
 	\def\theequation{\thesection.\arabic{equation}}}
\makeatother

\numberbysection

\newcommand{\be}{\begin{eqnarray}}
\newcommand{\ee}{\end{eqnarray}}
\newcommand{\non}{\nonumber}
\newcommand{\tr}{\mathop{\rm tr}\nolimits}
\newcommand{\id}{\mathbb{I}}
\newcommand{\csch}{\mathop{\rm csch}\nolimits}
\newcommand{\sech}{\mathop{\rm sech}\nolimits}
\newcommand{\M}{\mathop{\cal M}\nolimits}

\begin{document}

\begin{titlepage}
\strut\hfill UMTG--247
\vspace{.5in}
\begin{center}

\LARGE Generalized $T-Q$ relations\\
\LARGE and the open XXZ chain \\[1.0in]
\large Rajan Murgan and Rafael I. Nepomechie\\[0.8in]
\large Physics Department, P.O. Box 248046, University of Miami\\[0.2in]  
\large Coral Gables, FL 33124 USA\\

\end{center}

\vspace{.5in}

\begin{abstract}
We propose a generalization of the Baxter $T-Q$ relation which
involves more than one independent $Q(u)$.  We argue that the
eigenvalues of the transfer matrix of the open XXZ quantum spin chain
are given by such generalized $T-Q$ relations, for the case that at
most two of the boundary parameters $\{ \alpha_{-}, \alpha_{+},
\beta_{-}, \beta_{+} \}$ are nonzero, and the bulk anisotropy
parameter has values $\eta = {i \pi\over 2}\,, {i \pi\over 4}\,,
\ldots $.
\end{abstract}
\end{titlepage}

\setcounter{footnote}{0}

\section{Introduction}\label{sec:intro}

The famous Baxter $T-Q$ relation \cite{Ba}, which schematically
has the form
\be
t(u)\ Q(u) = Q(u') + Q(u'') \,,
\ee
holds for many integrable models associated with the $sl_{2}$ Lie
algebra and its deformations, such as the closed XXZ quantum spin chain.
This relation provides one of the most direct routes to the Bethe Ansatz
expression for the eigenvalues of the transfer matrix $t(u)$.

We propose here a generalization of this relation which involves more
than one $Q(u)$,
\be
t(u)\ Q_{1}(u) &=& Q_{2}(u') + Q_{2}(u'') \,, \non \\
t(u)\ Q_{2}(u) &=& Q_{1}(u''') + Q_{1}(u'''') \,.
\ee
This structure arises naturally in the open XXZ quantum spin chain
for special values of the bulk and boundary parameters.
We expect that such generalized $T-Q$ relations, involving two or more
independent $Q(u)$'s, may also appear in other integrable models.

The open XXZ chain with general integrable boundary terms 
\cite{dVGR} has a Hamiltonian which can be written as 
\be
{\cal H }&=& \sum_{n=1}^{N-1} H_{n\,, n+1} 
+{1\over 2}\sinh \eta \Big[ 
\coth \alpha_{-} \tanh \beta_{-}\sigma_{1}^{z}
+ \csch \alpha_{-} \sech \beta_{-}\big( 
\cosh \theta_{-}\sigma_{1}^{x} 
+ i\sinh \theta_{-}\sigma_{1}^{y} \big) \non \\
&-& \coth \alpha_{+} \tanh \beta_{+} \sigma_{N}^{z}
+ \csch \alpha_{+} \sech \beta_{+}\big( 
\cosh \theta_{+}\sigma_{N}^{x}
+ i\sinh \theta_{+}\sigma_{N}^{y} \big)
\Big] \,,
\label{Hamiltonian}
\ee
where $H_{n\,, n+1}$ is given by
\be
H_{n\,, n+1} = {1\over 2}\left( \sigma^{x}_{n}\sigma^{x}_{n+1}
+\sigma^{y}_{n}\sigma^{y}_{n+1}+\cosh \eta\ \sigma^{z}_{n}\sigma^{z}_{n+1}
\right) \,,
\label{twosite}
\ee
$\sigma^{x} \,, \sigma^{y} \,, \sigma^{z}$ are the usual Pauli
matrices, $\eta$ is the bulk anisotropy parameter, $\alpha_{\pm} \,,
\beta_{\pm} \,, \theta_{\pm}$ are boundary parameters, and $N$
is the number of spins. 

Although this model remains unsolved, the case of diagonal boundary terms
($\alpha_{\pm}$ or $\beta_{\pm} \rightarrow \pm \infty$)
was solved long ago \cite{Ga, ABBBQ, Sk}. Moreover, the case
of nondiagonal boundary terms when the boundary parameters obey
the constraint
\be
\alpha_{-} + \beta_{-} + \alpha_{+} + \beta_{+} = \pm (\theta_{-} - 
\theta_{+}) + \eta k \,,
\label{constraint}
\ee
(where $k \in \left[ -(N+1) \,, N+1 \right]$ is an even/odd integer if $N$ 
is odd/even, respectively) has recently been solved by
two approaches: generalized algebraic Bethe Ansatz \cite{CLSW},
and functional relations at roots of unity \cite{XXZ, Ne, NR}. 

It would clearly be desirable to overcome this constraint, and find
the solution for further values of the boundary parameters.  Some
progress was achieved recently using the functional relations approach
\cite{MN}.  Namely, Bethe Ans\"atze were proposed for the special cases
that at most one of the boundary parameters is nonzero, and the bulk
anisotropy has values $\eta = {i \pi\over 3}\,, {i \pi\over 5}\,,
\ldots $.

We find here (again by means of the functional relations approach)
that by allowing the possibility of generalized $T-Q$ relations, we
can obtain Bethe-Ansatz-type expressions for the transfer matrix
eigenvalues for the cases that at most {\it two} of the boundary parameters $\{
\alpha_{-}, \alpha_{+}, \beta_{-}, \beta_{+} \}$ are nonzero, and the bulk
anisotropy has values $\eta = {i \pi\over 2}\,, {i \pi\over 4}\,,
\ldots $.

In order to make the paper self-contained, we summarize in Section
\ref{sec:transfer} the construction of the transfer matrix and the
functional relations which it satisfies at roots of unity.
In order to derive the generalized $T-Q$ relation, it is instructive
to first understand why we are unable to obtain a conventional
relation with a single $Q(u)$. We present this analysis in Section 
\ref{sec:fail}. Finally, in Section \ref{sec:TQ} we derive the generalized 
$T-Q$ relations. We conclude in Section \ref{sec:discuss} with a 
brief discussion of our results and of some remaining open problems.

\section{Transfer matrix and functional relations at roots of unity}\label{sec:transfer}

The transfer matrix $t(u)$ of the open XXZ chain with general integrable
boundary terms, which satisfies the fundamental commutativity property
\be
\left[ t(u)\,, t(v) \right] = 0  \,,
\label{commutativity}
\ee 
is given by \cite{Sk}
\be
t(u) = \tr_{0} K^{+}_{0}(u)\  
T_{0}(u)\  K^{-}_{0}(u)\ \hat T_{0}(u)\,,
\label{transfer}
\ee
where $T_{0}(u)$ and $\hat T_{0}(u)$ are the monodromy matrices 
\be
T_{0}(u) = R_{0N}(u) \cdots  R_{01}(u) \,,  \qquad 
\hat T_{0}(u) = R_{01}(u) \cdots  R_{0N}(u) \,,
\label{monodromy}
\ee
and $\tr_{0}$ denotes trace over the ``auxiliary space'' 0.
The $R$ matrix is given by
\be
R(u) = \left( \begin{array}{cccc}
	\sinh  (u + \eta) &0            &0           &0            \\
	0                 &\sinh  u     &\sinh \eta  &0            \\
	0                 &\sinh \eta   &\sinh  u    &0            \\
	0                 &0            &0           &\sinh  (u + \eta)
\end{array} \right) \,,
\label{bulkRmatrix}
\ee 
where $\eta$ is the bulk anisotropy parameter; and $K^{\mp}(u)$ are
$2 \times 2$ matrices whose components
are given by \cite{dVGR, GZ}
\be
K_{11}^{-}(u) &=& 2 \left( \sinh \alpha_{-} \cosh \beta_{-} \cosh u +
\cosh \alpha_{-} \sinh \beta_{-} \sinh u \right) \non \\
K_{22}^{-}(u) &=& 2 \left( \sinh \alpha_{-} \cosh \beta_{-} \cosh u -
\cosh \alpha_{-} \sinh \beta_{-} \sinh u \right) \non \\
K_{12}^{-}(u) &=& e^{\theta_{-}} \sinh  2u \,, \qquad 
K_{21}^{-}(u) = e^{-\theta_{-}} \sinh  2u \,,
\label{Kminuscomponents}
\ee
and
\be
K_{11}^{+}(u) &=& -2 \left( \sinh \alpha_{+} \cosh \beta_{+} \cosh (u+\eta) 
- \cosh \alpha_{+} \sinh \beta_{+} \sinh (u+\eta) \right) \non \\
K_{22}^{+}(u) &=& -2 \left( \sinh \alpha_{+} \cosh \beta_{+} \cosh (u+\eta) 
+ \cosh \alpha_{+} \sinh \beta_{+} \sinh (u+\eta) \right) \non \\
K_{12}^{+}(u) &=& -e^{\theta_{+}} \sinh  2(u+\eta) \,, \qquad 
K_{21}^{+}(u) = -e^{-\theta_{+}} \sinh  2(u+\eta) \,,
\label{Kpluscomponents}
\ee
where $\alpha_{\mp} \,, \beta_{\mp} \,, \theta_{\mp}$ are the boundary
parameters. The transfer matrix ``contains'' the Hamiltonian (\ref{Hamiltonian}),
\be
{\cal H} = c_{1} {\partial \over \partial u} t(u) \Big\vert_{u=0} 
+ c_{2} \id \,,
\label{firstderivative}
\ee
where
\be
c_{1} &=& -{1\over 16 \sinh \alpha_{-} \cosh \beta_{-}
\sinh \alpha_{+} \cosh \beta_{+} \sinh^{2N-1} \eta 
\cosh \eta} \,, \non \\
c_{2} &=& - {\sinh^{2}\eta  + N \cosh^{2}\eta\over 2 \cosh \eta} 
\,,
\label{cees}
\ee 
and $\id$ is the identity matrix.

The transfer matrix eigenvalues $\Lambda(u)$ have $i\pi$ periodicity
\be
\Lambda(u+ i \pi) = \Lambda(u) \,,
\label{periodicity}
\ee
crossing symmetry
\be
\Lambda(-u - \eta)= \Lambda(u) \,,
\label{crossing}
\ee
and the asymptotic behavior 
\be
\Lambda(u) \sim -\cosh(\theta_{-}-\theta_{+})
{e^{u(2N+4)+\eta (N+2)}\over 2^{2N+1}} + 
\ldots \qquad \mbox{for} \qquad
u\rightarrow \infty \,.
\label{asymptotic}
\ee

For bulk anisotropy values 
\be
\eta = {i \pi\over p+1} \,, \qquad p= 1 \,, 2 \,, \ldots \,,
\label{eta}
\ee 
so that $q \equiv  e^{\eta}$ is a root of unity, the eigenvalues obey functional
relations of order $p+1$  \cite{XXZ, Ne}
\be
\lefteqn{\Lambda(u) \Lambda(u +\eta) \ldots \Lambda(u + p \eta)} \non \\
&-& \delta (u) \Lambda(u +2\eta) \Lambda(u +3\eta)
\ldots \Lambda(u + p \eta) \non \\
&-& \delta (u+\eta) \Lambda(u) \Lambda(u +3\eta) \Lambda(u +4\eta) 
\ldots \Lambda(u + p \eta) \non \\
&-& \delta (u+2\eta) \Lambda(u) \Lambda(u +\eta) \Lambda(u +4\eta) 
\ldots \Lambda(u + p \eta) - \ldots \non \\
&-& \delta (u+p\eta) \Lambda(u +\eta) \Lambda(u +2\eta) 
\ldots \Lambda(u + (p-1)\eta) \non \\
&+& \ldots  = f(u) \,.
\label{funcrltn}
\ee 
For example, for the case $p=3$, the functional relation is
\be
& &\Lambda(u) \Lambda(u+\eta) \Lambda(u+2\eta) \Lambda(u+3\eta) 
- \delta(u) \Lambda(u+2\eta) \Lambda(u+3\eta) 
- \delta(u+\eta) \Lambda(u) \Lambda(u+3\eta) \non \\
& &\quad  -\delta(u+2\eta) \Lambda(u) \Lambda(u+\eta) 
- \delta(u+3\eta) \Lambda(u+\eta) \Lambda(u+2\eta) \non \\
& &\quad +\delta(u) \delta(u+2\eta)
+ \delta(u+\eta) \delta(u+3\eta)
= f(u) \,.
\ee
The functions $\delta(u)$ and $f(u)$ are given in terms of the
boundary parameters  $\alpha_{\mp} \,, \beta_{\mp} \,, \theta_{\mp}$
by
\be
\delta(u) = \delta_{0}(u) \delta_{1}(u) \,, \qquad 
f(u) = f_{0}(u) f_{1}(u) \,,
\label{deltaf}
\ee
where
\be
\delta_{0}(u) &=& \left( \sinh u \sinh(u + 2\eta) \right)^{2N} {\sinh 2u
\sinh (2u + 4\eta)\over \sinh(2u+\eta) \sinh(2u+3\eta)}\,, \label{delta0} \\
\delta_{1}(u) &=&  2^{4} \sinh(u + \eta + \alpha_{-}) \sinh(u + \eta - \alpha_{-})
\cosh(u + \eta + \beta_{-}) \cosh(u + \eta - \beta_{-})  \non \\
& \times & \sinh(u + \eta + \alpha_{+}) \sinh(u + \eta - \alpha_{+})
\cosh(u + \eta + \beta_{+}) \cosh(u + \eta - \beta_{+}) \,,
\label{delta1}
\ee
and therefore, 
\be
\delta(u+ i\pi) =\delta(u) \,, \qquad \delta(-u -2\eta) =\delta(u)
\label{deltaproperties}
\,.
\ee 
For $p$ odd,
\be
f_{0}(u) &=& (-1)^{N+1} 2^{-2 p N} \sinh^{2N} \left( (p+1)u \right)
\tanh^{2} \left( (p+1)u \right)
\,, \label{f0odd} \\
f_{1}(u) &=& -2^{3-2 p} \Big( \non \\
& & \hspace{-0.2in}
\cosh \left( (p+1) \alpha_{-} \right)\cosh \left( (p+1) \beta_{-} \right)
\cosh \left( (p+1) \alpha_{+} \right)\cosh \left( (p+1) \beta_{+} \right)
\sinh^{2} \left( (p+1)u \right) \non \\
&-&
\sinh \left( (p+1) \alpha_{-} \right)\sinh \left( (p+1) \beta_{-} \right)
\sinh \left( (p+1) \alpha_{+} \right)\sinh \left( (p+1) \beta_{+} \right)
\cosh^{2} \left( (p+1)u \right) \non \\
&+&
(-1)^{N} \cosh \left( (p+1)(\theta_{-}-\theta_{+}) \right)
\sinh^{2} \left( (p+1)u \right) \cosh^{2} \left( (p+1)u \right) 
\Big) \,. \label{f1odd}
\ee 
For $p$ even,
\be
f_{0}(u) &=& (-1)^{N+1} 2^{-2 p N} \sinh^{2N} \left( (p+1)u \right)
\,, \label{f0} \\
f_{1}(u) &=& (-1)^{N+1} 2^{3-2 p} \Big( \non \\
& & \hspace{-0.2in}
\sinh \left( (p+1) \alpha_{-} \right)\cosh \left( (p+1) \beta_{-} \right)
\sinh \left( (p+1) \alpha_{+} \right)\cosh \left( (p+1) \beta_{+} \right)
\cosh^{2} \left( (p+1)u \right) \non \\
&-&
\cosh \left( (p+1) \alpha_{-} \right)\sinh \left( (p+1) \beta_{-} \right)
\cosh \left( (p+1) \alpha_{+} \right)\sinh \left( (p+1) \beta_{+} \right)
\sinh^{2} \left( (p+1)u \right) \non \\
&-&
(-1)^{N} \cosh \left( (p+1)(\theta_{-}-\theta_{+}) \right)
\sinh^{2} \left( (p+1)u \right) \cosh^{2} \left( (p+1)u \right) 
\Big) \,.
\label{f1}
\ee
Hence, $f(u)$ satisfies
\be
f(u + \eta) = f(u) \,, \qquad f(-u)=f(u) \,.
\ee
We also note the identity
\be
f_{0}(u)^{2} = \prod_{j=0}^{p}\delta_{0}(u + j \eta) \,.
\label{identity}
\ee

\section{An attempt to obtain a conventional $T-Q$ relation}\label{sec:fail}

In order to obtain Bethe Ansatz expressions for the transfer matrix
eigenvalues, we try (following \cite{BR}) to recast the functional
relations as the condition that the determinant of a certain matrix
vanishes. To this end, let us consider again the $(p+1) \times (p+1)$ matrix 
given by \cite{Ne}
\be
\M(u) = 
\left(
\begin{array}{cccccccc}
    \Lambda(u) & -{\delta(u)\over h(u+\eta)} & 0  & \ldots  & 0 & -h(u)  \\
    -h(u+\eta) & \Lambda(u+\eta) & -{\delta(u+\eta)\over h(u+2\eta)}  & \ldots  & 0 & 0  \\
    \vdots  & \vdots & \vdots & \ddots 
    & \vdots  & \vdots    \\
   -{\delta(u-\eta)\over h(u)}  & 0 & 0 & \ldots  & -h(u+p\eta) &
    \Lambda(u+p\eta)
\end{array} \right)  \,,
\label{oldM}
\ee
where $h(u)$ is a function which is
$i\pi$-periodic, but otherwise not yet specified.  Evidently, successive
rows of this matrix are obtained by simultaneously shifting $u \mapsto u+ \eta$ and
cyclically permuting the columns to the right.  Hence, this matrix has
the symmetry property
\be
S \M(u) S^{-1} = {\cal M}(u+\eta) \,,
\label{symmetryS}
\ee
where $S$ is the $(p+1) \times (p+1)$ matrix given by
\be
S = \left(
\begin{array}{cccccccc}
    0 & 1 & 0  & \ldots  & 0 & 0  \\
    0 & 0 & 1  & \ldots  & 0 & 0  \\
    \vdots  & \vdots & \vdots & \ddots 
    & \vdots  & \vdots    \\
    0 & 0 & 0  & \ldots  & 0 & 1 \\
   1  & 0 & 0 & \ldots  & 0 & 0
\end{array} \right) \,, \qquad S^{p+1} = 1 \,.
\label{Smatrix}
\ee

This symmetry implies that the corresponding $T-Q$
relation would involve only one $Q(u)$.
Indeed, if we assume $\det \M(u) =0$ (which, as we discuss below, turns out to be false
for the cases which we consider here), then $\M(u)$ has a null
eigenvector,
\be
\M(u)\ v(u) = 0 \,.
\label{nulleigenvector}
\ee 
The symmetry (\ref{symmetryS}) is consistent with
\be
S\ v(u) = v(u+\eta) \,,
\ee
which in turn implies that $v(u)$ has the form 
\be
v(u) = \left( Q(u)\,, Q(u+\eta) \,, \ldots \,, Q(u+p\eta) \right) 
\,, \qquad Q(u+i\pi) = Q(u) \,.
\label{oldv}
\ee
That is, all the components of $v(u)$ are determined by a single
function $Q(u)$. The null eigenvector condition (\ref{nulleigenvector})
together with the explicit forms  (\ref{oldM}),  (\ref{oldv}) of $\M(u)$ and $v(u)$
would then lead to a conventional $T-Q$ relation.

One can verify that the condition $\det \M(u) =0$ indeed implies the
functional relations (\ref{funcrltn}), if $h(u)$ satisfies
\be
f(u) = \prod_{j=0}^{p} h(u+j\eta) 
+ \prod_{j=0}^{p}{ \delta(u+j\eta)\over h(u+j\eta)}  \,.
\label{conditiononh}
\ee
Setting 
\be
z(u) \equiv \prod_{j=0}^{p} h(u+j\eta) \,, 
\label{zdefinition}
\ee
it immediately follows from (\ref{conditiononh}) that $z(u)$ is given by
\be 
z(u) = {1\over 2}\left( f(u) \pm \sqrt{\Delta(u)} \right) \,, 
\label{zresult}
\ee 
where $\Delta(u)$ is defined by 
\be
\Delta(u) \equiv  f(u)^{2} -4  \prod_{j=0}^{p} \delta(u+j\eta) \,.
\label{Delta}
\ee 

We wish to focus here on new special cases that $\Delta(u)$ is a perfect square.
\footnote{When the constraint (\ref{constraint}) is satisfied, $\Delta(u)$ 
is a perfect square; these are the cases studied in \cite{XXZ}.
For even values of $p$, $\Delta(u)$ is also a perfect square if at 
most one of the boundary parameters is nonzero; these are the cases studied in \cite{MN}.}
For odd values of $p$, $\Delta(u)$ is also a perfect square if 
at most {\it two} of the boundary parameters $\{
\alpha_{-}, \alpha_{+}, \beta_{-}, \beta_{+} \}$ are nonzero. 
{\bf We henceforth restrict to such parameter values.} In particular,
we assume that $\eta$ is given by (\ref{eta}), with $p$ odd (i.e., 
bulk anisotropy values $\eta = {i\pi\over 2}\,, {i \pi\over 4}\,, \ldots $).
For definiteness, here we present results for the case that $\alpha_{-},
\beta_{-} \ne 0$ and $\alpha_{+}=\beta_{+}=\theta_{\pm}=0$. (In the Appendix, we 
present results for the case that $\alpha_{\pm} \ne 0$ and
$\beta_{\pm}=\theta_{\pm}=0$; and similar results hold for the other cases.)
Moreover, we also restrict to even values of $N$. (We expect similar results 
to hold for odd $N$.)

For such parameter values, it is easy to arrive at a contradiction.
Indeed, on one hand, the definition (\ref{zdefinition}) together with the assumed
$i\pi$-periodicity of $h(u)$ (which is required for the symmetry
(\ref{symmetryS})) imply the result $z(u) = z(u+\eta)$. On the other hand,
(\ref{Delta}) together with (\ref{deltaf})-(\ref{f1odd}) and
(\ref{identity}) imply
\be
\sqrt{\Delta(u)} = 2^{3-2p} f_{0}(u) \left(\cosh((p+1)\alpha_{-}) +
\cosh((p+1)\beta_{-}) \right) \sinh^{2}((p+1)u) \cosh ((p+1)u) \,.
\label{explicitDelta}
\ee
Hence, it follows from (\ref{zresult}) that $z(u) \ne z(u+\eta)$,
which contradicts the earlier result.
We conclude that for such parameter values, it is {\it not} possible to 
find a function $h(u)$ which is $i\pi$-periodic and 
satisfies the condition (\ref{conditiononh}).
Hence, for such parameter values,
the matrix $\M(u)$ given by (\ref{oldM}) does {\it not} lead to the solution of the
model, and we fail to obtain a conventional $T-Q$ relation.

We remark that if either $\alpha_{+}$ or $\alpha_{-}$ is zero, then
the Hamiltonian is no longer given by (\ref{Hamiltonian}), since the
coefficient $c_{1}$ (\ref{cees}) is singular. Indeed, as noted in \cite{MN},
$t'(0)$ is then proportional to $\sigma^{x}_{N}$. Hence, in order
to obtain a nontrivial integrable Hamiltonian, one must consider the second
derivative of the transfer matrix. For the case $\alpha_{-} \,,
\beta_{-} \ne 0$,
\be
t''(0) &=& -16 \sinh^{2N-1} \eta \cosh \eta 
\Bigg(  \sinh \alpha_{-} \cosh \beta_{-}\left\{ 
\sigma^{x}_{N} \,, \sum_{n=1}^{N-1} H_{n\,, n+1} \right\} \non \\
&+& \sinh \alpha_{-} \cosh \beta_{-} (N \cosh \eta + \sinh \eta \tanh \eta) 
\sigma^{x}_{N} \non \\
&+& \sinh \eta \left(\sigma^{x}_{1} 
+ \sinh \beta_{-} \cosh \alpha_{-} \sigma^{z}_{1} \right) \sigma^{x}_{N} 
\Bigg)
\,,
\ee 
where $H_{n\,, n+1}$ is given by (\ref{twosite}). The case
$\alpha_{\pm}\ne 0$, for which the Hamiltonian instead has a conventional
local form, is discussed in the Appendix.

\section{The generalized $T-Q$ relations}\label{sec:TQ}

Instead of demanding the symmetry (\ref{symmetryS}), let us now demand 
only the weaker symmetry
\be
T \M(u) T^{-1} = {\cal M}(u+2\eta) \,, \qquad T \equiv S^{2} \,,
\label{symmetryT}
\ee
where $S$ is given by (\ref{Smatrix}).
Indeed, (\ref{symmetryS}) implies  (\ref{symmetryT}), but the 
converse is not true. A matrix $\M(u)$ with such symmetry is given by
\be
\M(u) = 
\left(
\begin{array}{cccccccc}
    \Lambda(u) & -{\delta(u)\over h^{(1)}(u)} & 0  & \ldots  & 0 &
    -{\delta(u-\eta)\over h^{(2)}(u-\eta)}  \\
    -h^{(1)}(u) & \Lambda(u+\eta) & -h^{(2)}(u+\eta)  & \ldots  & 0 & 0  \\
    \vdots  & \vdots & \vdots & \ddots 
    & \vdots  & \vdots    \\
   -h^{(2)}(u-\eta)  & 0 & 0 & \ldots  & -h^{(1)}(u+(p-1)\eta) &
    \Lambda(u+p\eta) 
\end{array} \right)  \,,
\label{newM}
\ee
where $h^{(1)}(u)$ and $h^{(2)}(u)$ are functions which are
$i\pi$-periodic, but otherwise not yet specified.

This symmetry implies that the corresponding $T-Q$ relations will 
involve {\it two} $Q(u)$'s. Indeed, assuming again that 
\be
\det \M(u) =0 \,,
\label{vanishingdet}
\ee 
then $\M(u)$ has a null eigenvector $v(u)$,
\be
\M(u)\ v(u) = 0 \,.
\label{newnulleigenvector}
\ee 
The symmetry (\ref{symmetryT}) is consistent with
\be
T\ v(u) = v(u+2\eta) \,,
\ee
which implies that $v(u)$ has the form 
\be
v(u) = \left( Q_{1}(u)\,, Q_{2}(u) \,, \ldots \,, Q_{1}(u-2\eta) 
\,, Q_{2}(u-2\eta)\right) \,, 
\label{newv} 
\ee
with
\be 
Q_{1}(u) = Q_{1}(u+i\pi) \,, \qquad  Q_{2}(u) = Q_{2}(u+i\pi) \,.
\label{Qperiodicity}
\ee
That is, the components of $v(u)$ are determined by {\it two} 
independent 
functions, $Q_{1}(u)$ and $Q_{2}(u)$. The null eigenvector condition 
(\ref{newnulleigenvector})
together with the explicit forms  (\ref{newM}),  (\ref{newv}) of $\M(u)$ and $v(u)$
now lead to generalized $T-Q$ relations,
\be
\Lambda(u) &=& 
{\delta(u)\over h^{(1)}(u)} {Q_{2}(u)\over Q_{1}(u)} 
+ {\delta(u-\eta)\over h^{(2)}(u-\eta)} {Q_{2}(u-2\eta)\over 
Q_{1}(u)} \,, \label{TQ1} \\
 &=& 
h^{(1)}(u-\eta) {Q_{1}(u-\eta)\over Q_{2}(u-\eta)} 
+ h^{(2)}(u) {Q_{1}(u+\eta)\over 
Q_{2}(u-\eta)} \,.
\label{TQ2}
\ee

Since $\Lambda(u)$ has the crossing symmetry (\ref{crossing}) and $\delta(u)$ has 
the crossing property (\ref{deltaproperties}), it is natural to have 
the two terms in (\ref{TQ1}) transform into each other under crossing.
Hence, we set
\be
h^{(2)}(u) = h^{(1)}(-u-2\eta) \,,
\label{h2}
\ee
and we make the Ansatz
\be
Q_{1}(u) &=& \prod_{j=1}^{M_{1}} 
\sinh (u - u_{j}^{(1)}) \sinh (u + u_{j}^{(1)} + \eta) \,, \non \\
Q_{2}(u) &=& \prod_{j=1}^{M_{2}} 
\sinh (u - u_{j}^{(2)}) \sinh (u + u_{j}^{(2)} + 3\eta) \,,
\label{ansatz}
\ee 
which is consistent with the required periodicity (\ref{Qperiodicity})
and crossing properties
\be
Q_{1}(u) &=& Q_{1}(-u-\eta) \,, \qquad Q_{2}(u) = Q_{2}(-u-3\eta) \,.
\ee

Analyticity of $\Lambda(u)$ (\ref{TQ1}), (\ref{TQ2}) implies
Bethe-Ansatz-type equations for the zeros $\{ u_{j}^{(1)} \,,
u_{j}^{(2)} \}$ of $Q_{1}(u)\,, Q_{2}(u)$, respectively,
\be
{\delta(u_{j}^{(1)})\ h^{(2)}(u_{j}^{(1)}-\eta)
\over \delta(u_{j}^{(1)}-\eta)\ h^{(1)}(u_{j}^{(1)})} 
&=&-{Q_{2}(u_{j}^{(1)}-2\eta)\over Q_{2}(u_{j}^{(1)})} \,, \qquad j =
1\,, 2\,, \ldots \,, M_{1} \,, \non \\
{h^{(1)}(u_{j}^{(2)})\over h^{(2)}(u_{j}^{(2)}+\eta)}
&=&-{Q_{1}(u_{j}^{(2)}+2\eta)\over Q_{1}(u_{j}^{(2)})} \,, \qquad j =
1\,, 2\,, \ldots \,, M_{2} \,.
\label{BAE}
\ee 

Note that the function $h^{(1)}(u)$ has not yet been specified, nor
has the important assumption that $\M(u)$ has a vanishing determinant
(\ref{vanishingdet}) yet been verified.  These problems are
closely related, and we now address them both.

One can verify that the condition $\det \M(u) =0$ indeed implies the
functional relations (\ref{funcrltn}), if $h^{(1)}(u)$ satisfies
\be
f(u) = w(u) \prod_{j=0,2,\ldots}^{p-1} \delta(u+j\eta) 
+ {1\over w(u)} \prod_{j=1,3,\ldots}^{p} \delta(u+j\eta) \,,
\label{newconditiononh}
\ee
where 
\be
w(u) \equiv {\prod_{j=1,3,\ldots}^{p} h^{(2)}(u+j\eta)\over 
\prod_{j=0,2,\ldots}^{p-1} h^{(1)}(u+j\eta)} \,.
\label{wdefinition}
\ee
It immediately follows from (\ref{newconditiononh}) that $w(u)$ is given by
\be 
w(u) = {f(u) \pm \sqrt{\Delta(u)}\over 2 \prod_{j=0,2,\ldots}^{p-1} \delta(u+j\eta)} \,, 
\label{wresult}
\ee 
where $\Delta(u)$ is the same quantity defined in (\ref{Delta}).

Let us recall that we are considering the case that $p$ is odd, and 
that at most $\alpha_{-}$ and $\beta_{-}$ are nonzero. For this case,
$\sqrt{\Delta(u)}$ is given by (\ref{explicitDelta}). It follows 
from (\ref{wresult}) that for $p= 3\,, 7\,, 11\,, \ldots$ the
two solutions for $w(u)$ are given by
\be
w(u) &=& \coth^{2N}\left({1\over 2}(p+1) u\right) \,, \non \\
w(u) &=& \left({\cosh((p+1) u) - \cosh((p+1) \alpha_{-})\over
\cosh((p+1) u) + \cosh((p+1) \alpha_{-})} \right)
\left({\cosh((p+1) u) - \cosh((p+1) \beta_{-})\over
\cosh((p+1) u) + \cosh((p+1) \beta_{-})} \right) \non \\
& & \quad \times \coth^{2N}\left({1\over 2}(p+1) u\right) \,, \qquad
p= 3\,, 7\,, 11\,, \ldots \,;
\label{w1}
\ee 
and for $p= 1\,, 5\,, 9\,, \ldots$ the two solutions for $w(u)$ are given by
\be
w(u) &=& \left({\cosh((p+1) u) - \cosh((p+1) \alpha_{-})\over
\cosh((p+1) u) + \cosh((p+1) \alpha_{-})} \right)
\coth^{2N}\left({1\over 2}(p+1) u\right) \,, \non \\
w(u) &=& 
\left({\cosh((p+1) u) + \cosh((p+1) \beta_{-})\over
\cosh((p+1) u) - \cosh((p+1) \beta_{-})} \right) 
\coth^{2N}\left({1\over 2}(p+1) u\right) \,,  \non \\
& &\quad p= 1\,, 5\,, 9\,, \ldots \,.
\label{w2}
\ee 

There are many solutions of (\ref{wdefinition}) for $h^{(1)}(u)$ (with
$h^{(2)}(u)$ given by (\ref{h2})) corresponding to the above expressions 
for $w(u)$, which also have the required $i\pi$ periodicity.
We consider here the solutions
\be
h^{(1)}(u) = -4 \sinh^{2N}(u + 2\eta) \,, \quad 
M_{2} = {1\over 2} N + p - 1 \,, \quad M_{1} = M_{2} + 2  \,, \quad
p= 3\,, 7\,, 11\,, \ldots
\label{h11}
\ee
and 
\be
h^{(1)}(u) = \left\{ 
\begin{array}{ll}
    -2 \cosh(u+\alpha_{-})\cosh(u-\alpha_{-})\cosh(2u) \sinh^{2N}(u + 2\eta)\,, 
    & M_{1} =M_{2} = {1\over 2} N + 2p - 1 \,, \\
\qquad \qquad p= 9\,, 17\,, 25\,, \ldots \\
    2 \cosh(u+\alpha_{-})\cosh(u-\alpha_{-})\cosh(2u) \sinh^{2N}(u + 2\eta)\,, 
    & M_{1} =M_{2} = {1\over 2} N + {3\over 2}(p - 1) \,, \\
\qquad \qquad p= 5\,, 13\,, 21\,, \ldots \\
    2 \cosh(u+\alpha_{-})\cosh(u-\alpha_{-})\cosh(2u) \sinh^{2N}(u +
    2\eta)\,, & M_{1} =M_{2} = {1\over 2} N + 2 \,, \\
\qquad \qquad p=1 \,,
\end{array} \right.
\label{h12}
\ee
corresponding to the first solutions for $w(u)$ given in (\ref{w1}),  
(\ref{w2}), respectively. We have searched for
solutions largely by trial and error, verifying
numerically (along the lines explained in \cite{NR}) for small values 
of $N$ that the eigenvalues can indeed be expressed as 
(\ref{TQ1}), (\ref{TQ2}) with $Q(u)$'s of the form (\ref{ansatz}). 

Note that the values of $M_{1}$ and 
$M_{2}$ (i.e., the number of zeros of $Q_{1}(u)$ and $Q_{2}(u)$,
respectively) depend on the particular choice for the function $h^{(1)}(u)$. 
Our reason for choosing (\ref{h11}),  (\ref{h12}) 
over the other solutions which we found is that the former solutions gave the {\it 
lowest} values of $M_{1}$ and $M_{2}$, for given values of $N$ and $p$.
(It would be interesting to know whether there exist other solutions
for $h^{(1)}(u)$ which give even lower values of $M_{1}$ and $M_{2}$.)
Our conjectured values of $M_{1}$ and $M_{2}$ given in (\ref{h11}), 
(\ref{h12}) are consistent with 
the asymptotic behavior (\ref{asymptotic}). Moreover, these values have been checked
numerically for small values of $N$ (up to $N=6$) and $p$ (up to 
$p=21$). That is, we have verified 
numerically that, with the above choice of $h^{(1)}(u)$, the generalized 
$T-Q$ relations (\ref{TQ1}), (\ref{TQ2}) correctly give all $2^{N}$
eigenvalues, with
$Q_{1}(u)$ and $Q_{2}(u)$ of the form (\ref{ansatz}) and with $M_{1}$
and $M_{2}$ given in (\ref{h11}), (\ref{h12}). We expect that similar results 
can be obtained corresponding to the second solutions for $w(u)$.

To summarize, we propose that for the case that $p$ is odd and 
that at most $\alpha_{-}$, $\beta_{-}$ are nonzero,
the eigenvalues $\Lambda(u)$ of the transfer matrix $t(u)$ 
(\ref{transfer}) are given by the generalized 
$T-Q$ relations (\ref{TQ1}), (\ref{TQ2}), with 
$Q_{1}(u)$ and $Q_{2}(u)$ given by (\ref{ansatz}), $h^{(2)}(u)$ given 
by  (\ref{h2}), and $h^{(1)}(u)$ given 
by (\ref{h11}), (\ref{h12}). The zeros $\{ u_{j}^{(1)} \,, u_{j}^{(2)} \}$ of
$Q_{1}(u)$ and $Q_{2}(u)$ are solutions of the Bethe Ansatz equations (\ref{BAE}).
We expect that there are sufficiently many such equations to determine all
the zeros. As already mentioned, similar results hold for the case
that at most two of the boundary parameters $\{
\alpha_{-}, \alpha_{+}, \beta_{-}, \beta_{+} \}$ are nonzero.

\section{Discussion}\label{sec:discuss}

We have argued that the eigenvalues of the transfer matrix of the open
XXZ chain, for the special case that $p$ is odd and that at most two of
the boundary parameters $\{ \alpha_{-}, \alpha_{+}, \beta_{-},
\beta_{+} \}$ are nonzero, can be given by generalized 
$T-Q$ relations (\ref{TQ1}), (\ref{TQ2}) involving more than one 
$Q(u)$. Although we have not ruled out the possibility of expressing these 
eigenvalues in terms of a conventional $T-Q$ relation, the analysis 
in Section \ref{sec:fail} suggests to us that this is unlikely.

Many interesting problems remain to be explored. It should be possible 
to explicitly construct {\it operators} $Q_{1}(u)$, $Q_{2}(u)$ which commute 
with each other and with the transfer matrix $t(u)$, and whose 
eigenvalues are given by (\ref{ansatz}). There may be 
further special cases for which the quantity $\Delta(u)$ (\ref{Delta}) is a 
perfect square, in which case it should not be difficult to find the 
corresponding Bethe Ansatz solution. The general case that $\Delta(u)$ is 
{\it not} a perfect square and/or that $\eta \ne i\pi/(p+1)$ remains to be 
understood.

Generalized $T-Q$ relations are novel structures, which merit further
investigation.  The corresponding Bethe Ansatz equations (e.g.,
(\ref{BAE})) have some resemblance to the ``nested'' equations which
are characteristic of higher-rank models.  Such generalized $T-Q$
relations, involving two or even more $Q(u)$'s, may also lead to
further solutions of integrable open chains of higher rank and/or
higher-dimensional representations.  (For recent progress on such
models, see e.g. \cite{generalization}.)

\section*{Acknowledgments}

One of us (RN) is grateful to S. Ruijsenaars for helpful
correspondence.
This work was supported in part by the National Science Foundation
under Grant PHY-0244261.

\appendix

\section{Appendix: the case $\alpha_{\pm} \ne 0$ and
$\beta_{\pm}=\theta_{\pm}=0$}

Here we consider the case that $\alpha_{\pm} \ne 0$ and
$\beta_{\pm}=\theta_{\pm}=0$, for which the Hamiltonian is local, 
\be
{\cal H }&=& \sum_{n=1}^{N-1} H_{n\,, n+1} 
+{1\over 2}\sinh \eta \Big( 
\csch \alpha_{-} \sigma_{1}^{x} 
+ \csch \alpha_{+} \sigma_{N}^{x} \Big) \,,
\label{niceHamiltonian}
\ee
as follows from (\ref{Hamiltonian}).
For this case, the quantity $\sqrt{\Delta(u)}$ is given by (\ref{explicitDelta}) 
with $\beta_{-}$ replaced by $\alpha_{+}$, namely, 
\be
\sqrt{\Delta(u)} = 2^{3-2p} f_{0}(u) \left(\cosh((p+1)\alpha_{-}) +
\cosh((p+1)\alpha_{+}) \right) \sinh^{2}((p+1)u) \cosh ((p+1)u) \,.
\ee
It follows that the two solutions for $w(u)$ (\ref{wresult}) are given by
\be
w(u) &=& \coth^{2N}\left({1\over 2}(p+1) u\right) \,, \non \\
w(u) &=& \left({\cosh((p+1) u) - \cosh((p+1) \alpha_{-})\over
\cosh((p+1) u) + \cosh((p+1) \alpha_{-})} \right)
\left({\cosh((p+1) u) - \cosh((p+1) \alpha_{+})\over
\cosh((p+1) u) + \cosh((p+1) \alpha_{+})} \right) \non \\
& & \quad \times \coth^{2N}\left({1\over 2}(p+1) u\right) \,, \qquad
p= 3\,, 7\,, 11\,, \ldots \,,
\label{w1app}
\ee 
and
\be
w(u) &=& 
\coth^{2N+2}\left({1\over 2}(p+1) u\right) \,, \non \\
w(u) &=&\left({\cosh((p+1) u) - \cosh((p+1) \alpha_{-})\over
\cosh((p+1) u) + \cosh((p+1) \alpha_{-})} \right)
\left({\cosh((p+1) u) - \cosh((p+1) \alpha_{+})\over
\cosh((p+1) u) + \cosh((p+1) \alpha_{+})} \right) \non \\
& & \quad \times \coth^{2N+2}\left({1\over 2}(p+1) u\right) \,, \qquad
p= 1\,, 5\,, 9\,, \ldots \,.
\label{w2app}
\ee 
For simplicity, let us once again consider just the first solutions 
for $w(u)$ given in (\ref{w1app}) and (\ref{w2app}), which are
independent of $\alpha_{\pm}$. Corresponding 
solutions of (\ref{wdefinition}) for $h^{(1)}(u)$ (with
$h^{(2)}(u)$ given by (\ref{h2})) are
\be
h^{(1)}(u) = 4 \sinh^{2N}(u + 2\eta) \,, \quad 
M_{2} = {1\over 2} N + {1\over 2}(3p - 1) \,, \quad M_{1} = M_{2} + 2  \,, \quad
p= 3\,, 7\,, 11\,, \ldots
\label{h11app}
\ee
and
\be
h^{(1)}(u) = \left\{ 
\begin{array}{ll}
    -2\cosh(2u) \sinh^{2}u \sinh^{2N}(u + 2\eta)\,, 
    & M_{1} =M_{2} = {1\over 2} N + 2p - 1 \,, \\
\qquad \qquad p= 9\,, 17\,, 25\,, \ldots \\
   2\cosh(2u) \sinh^{2}u \sinh^{2N}(u + 2\eta)\,,  
    & M_{1} =M_{2} = {1\over 2} N + {3\over 2}(p - 1) \,, \\
\qquad \qquad p= 5\,, 13\,, 21\,, \ldots \\
   2\cosh(2u) \sinh^{2}u  \sinh^{2N}(u +
    2\eta)\,, & M_{1} =M_{2} = {1\over 2} N + 2 \,, \\
\qquad \qquad p=1 \,.
\end{array} \right.
\label{h12app}
\ee
That is, the eigenvalues $\Lambda(u)$ of the transfer matrix $t(u)$ 
(\ref{transfer}), for $\eta$ values (\ref{eta}) with $p$ odd and for
$\alpha_{\pm} \ne 0$ and
$\beta_{\pm}=\theta_{\pm}=0$, are given by the generalized 
$T-Q$ relations (\ref{TQ1}), (\ref{TQ2}), with 
$Q_{1}(u)$ and $Q_{2}(u)$ given by (\ref{ansatz}), $h^{(2)}(u)$ given 
by  (\ref{h2}), and $h^{(1)}(u)$ given 
by (\ref{h11app}), (\ref{h12app}). The zeros $\{ u_{j}^{(1)} \,, u_{j}^{(2)} \}$ of
$Q_{1}(u)$ and $Q_{2}(u)$ are solutions of the Bethe Ansatz equations (\ref{BAE}).

\end{document}